\documentclass[superscriptaddress,10pt]{revtex4}

\newcommand{\TM}{(TM)}
\newcommand{\Intel}{ Intel\textregistered~ }
\newcommand{\OPA}{ \Intel Omni-Path\TM~}
\newcommand{\KNL}{ \Intel Knight's Landing\TM~}
\newcommand{\XeonPhi}{ \Intel Xeon Phi\TM Processor~}
\newcommand{\SKL}{ \Intel Scalable Family processor (Skylake)}

\newcommand{\Xeon}{ \Intel Xeon\textregistered~}
\usepackage{textcomp}
\usepackage{hyperref}
\usepackage{graphicx}
\usepackage{bm}
\usepackage{placeins}
\usepackage{multirow}
\usepackage{graphicx}
\usepackage{epsfig}
\usepackage{psfrag}
\usepackage{soul}
\usepackage{color}
\usepackage{multirow}
\usepackage{mathptmx}
\usepackage{mathrsfs}
\usepackage{amsmath, amssymb}
\usepackage{slashed}

\newcommand{\beq}{\begin{equation}}
\newcommand{\eeq}{\end{equation}}
\newcommand{\beqa}{\begin{eqnarray}}
\newcommand{\eeqa}{\end{eqnarray}}

\newcommand{\FslashA}[1]{\!\not{\hbox{\kern-2pt ${#1}$}}}
\newcommand{\FslashB}[1]{\!\not{\hbox{\kern+1pt ${#1}$}}}

\allowdisplaybreaks

\usepackage{dcolumn}

\begin{document}


\title{Accelerating HPC codes on \Intel Omni-Path Architecture networks: From particle physics to Machine Learning}

\author{Peter Boyle}\affiliation{The University of Edinburgh and Alan Turing Institute}
\author{Michael Chuvelev}\affiliation{Intel} 
\author{Guido Cossu}\affiliation{The University of Edinburgh}
\author{Christopher Kelly}\affiliation{Columbia University}
\author{Christoph Lehner}\affiliation{Brookhaven National Laboratory}
\author{Lawrence Meadows}\affiliation{Intel}

\begin{abstract}
We discuss practical methods to ensure near wirespeed performance from clusters
with either one or two \Intel Omni-Path host fabric interfaces (HFI) per node,
and \Intel Xeon Phi(TM) 72xx (Knight's Landing) processors, and using
the Linux operating system.

The study evaluates the performance improvements achievable and the required programming
approaches in two distinct example problems: firstly in Cartesian communicator halo exchange problems, appropriate for structured grid PDE solvers
that arise in quantum chromodynamics simulations of particle physics; and
secondly in gradient reduction appropriate to synchronous stochastic gradient descent for
machine learning. As an example, we accelerate a published Baidu Research reduction code and
obtain a factor of ten speedup over the original code using the techniques discussed in this paper.
This displays how a factor of ten speedup in strongly scaled distributed machine learning could be achieved when
synchronous stochastic gradient descent is massively parallelised with a fixed mini-batch size.

We find a significant improvement in performance robustness when memory
is obtained using carefully (guaranteed) allocated 2MB ``huge'' virtual memory pages, implying that non-standard
allocation routines should be used for communication buffers. These can be easily accessed via a LD\_PRELOAD override
in the manner suggested by libhugetlbfs, or accessed by appropiate mmap calls.
We make use of the \Intel MPI 2019 library ``Technology Preview'' and underlying software
to enable thread concurrency throughout the communication software stack via multiple PSM2
endpoints per process and use of multiple independent MPI communicators. 
When using a single MPI process per node, we find
that this greatly accelerates delivered bandwidth in many core \Intel Xeon Phi
processors.
\end{abstract}

\maketitle

\section{Introduction}

Modern massively parallel supercomputers are composed of many (relatively) commodity computing elements,
which may communicate using a variety of interconnect technologies which are usually programmed through
the now standard Message Passing Interface (MPI).
State of the art interconnect technologies are able to offload the work of copying 
memory resident data to and from the network, and enable \emph{zero copy direct memory access} 
where data is read from or deposited directly to user space buffers without any intermediate
copy to kernel or device driver memory.

This study evaluates the performance improvements achievable and the required programming
approaches: firstly in Cartesian communicator halo exchange problems, appropriate for structured grid PDE solvers; and
secondly in gradient reduction appropriate to synchronous stochastic gradient descent for
machine learning.
The systems under test will use either one or two \OPA\cite{OPApaper,OPAtalk} host fabric interfaces (HFIs) per node,
and \KNL \cite{Sodani} processors,
and use the Linux operating system which has become the dominant software
platform for High Performance Computing (HPC).

The structure of this paper is as follows: we will firstly discuss some background
computer architecture, explaining some underlying interconnect and operating system 
concepts at a basic level suitable to computational scientists
who are not experts in computer design; we will then 
present recommend techniques to introduce concurrent multithread reentrancy through the 
Intel MPI 2019 Technology Preview; we will also show that whether or not the new MPI library
is used, there is a large improvement in the stability of the performance afforded by the reliable use
of explicit huge memory page allocations due to the suppression of per-page software overhead
(which may depend on the page fragmentation history of a node prior to job execution; something beyond
the control of a user).

\section{Background}

In order to understand the reason for the improvements discussed in this paper, some
background knowledge of the handling of virtual memory and device access by modern operating systems is required,
and a brief summary is given in this section.

\subsection{Virtual memory and zero copy DMA}

Memory protection between processes is used to both abstract the physical memory map, protect
multiple processes from each other's accesses, and allow the sum of all processes memory to exceed
physical dram capacity via swapping mechanisms.
Memory in user programmes is \emph{virtually addressed}: the addresses used by software are
translated by hardware mechanisms in a memory management unit (MMU) 
on a page by page granularity; each load or store involves looking up a \emph{page table}
that resolves presence, access permissions, and the location in physical memory of each
virtual page the kernel offers to a user process. The physical and virtual address spaces are
kept completely distinct, and a user programme can never see, interpret or use a physical address.
The page size and hardware page table varies from archicture to architecture.
The default page size remains at 4KB in the x86 architecture, with architecture extensions
allowing Linux to allocate 2MB and 1GB in certain circumstances.

When a memory access to a virtual page is performed by a CPU core, the
hardware must first use the memory resident page table to discover the corresponding
physical page. If not present, or access permissions dictate, a page fault interrupt 
is triggered to invoke the kernel to handle the condition. 

We may place the penalties associated with virtual translation into two broad categories:
\begin{enumerate}
\item[A)] Those that invoke software overhead on a per page level granularity. 
\item[B)] Those that invoke a hardware state machine to \emph{walk} a memory hardware page table.
\end{enumerate}

The former include Unix page faults, and can cost several orders of magnitude
more to resolve than a TLB miss and hardware page table walk. 

A good example of page faults (A) (beyond the  \emph{segmentation violation} indicating attempted access to an invalid virtual address,
and familiar to most error prone programmers) 
is the mechanism used by GNU/Linux to handle large, new memory allocations.
The glibc malloc routines use
the \emph{mmap} system call to map an \emph{anonymous} region 
(i.e. process private and \emph{not} associated with a file) 
for large (multi-page) allocations. 
The Linux kernel is firstly invoked to add and remove the
virtual address pages from the user process during each allocate and free. 
Linux performs immediate virtual page allocation, but \emph{deferred} physical allocation to prevent potentially
needless and long latency operation by the kernel. Linux rather initially maps the \emph{same}
write protected ``zero'' physical page (containing exclusively zero data) multiple times throughout
the virtual address range. The software state of each virtual page is ``copy-on-write'',
such that when making the first write attempt to each page in the region, a \emph{page fault} is raised with 
the kernel and serviced by allocating a new physical page, and \emph{copying} the original
page to the new, write allowed, location and replaying the write instruction (which now succeeds to the new, write
enabled, physical page) on return from interrupt to the user process.

It is easy to imagine many forms of software overhead requiring the few microsecond timescale to resolve. 
A back of the envelope calculation
suggests processing 4KB pages at a 0.5 MHZ rate can attain at most 2GB/s data processing rate
when this overhead is paid on a per page basis. This mode for performance loss is a key 
aspect of the issues addressed by this paper.

For hardware translation events (B), 
the cost is only of order a few times the memory access latency.
Recent translations are cached in a translation lookaside buffer (TLB), so that multiple
references to the same page invoke only a single page table search and the overhead
of this hardware translation is typically small, since under typical usage a reasonable level of locality
of memory references and adequate capacity for caching these translations leaves only small fraction references
missing this translation cache. 
In recent x86 chips the level 1 TLB cache holds 64 entries, and the level 2 TLB cache contains 1536 entries.
With 4KB pages these cover a 256KB and 6MB working set respectively, and with 2MB pages they cover
8MB and 3GB respectively.
Database applications may present pathological cases, with
access to small records that are scattered all over the virtual memory space. It is not uncommon for such
applications to make use of the ability to reserve huge pages in order to avoid paying
the TLB miss penalty with nearly every memory access\cite{Oracle}. We will show in this paper
that similar strategies must now be adopted to obtain best performance with applications
attempting to use zero-copy DMA from user space with the \OPA interconnect.

\subsection{Zero copy DMA devices}

For devices, such as an interconnect adaptor, it is desirable to eliminate copying of data
by injecting or retrieving data directly from user process virtually addressed memory.
To achieve this, several steps must be taken in a multiprocessor computer.
To offload the work of a bulk copy, the device can provide a hardware \emph{direct memory
access} (DMA) engine.

However, in normal circumstances, the operating system makes no guarantees about the physical
location or even presence of the pages of a user process memory in the physical memory. 
They may be relocated or swapped to disk as dictated by the operating system, running concurrently
multiple cores, while it juggles multiple tasks responding to their competing demands 
for the physical resources of the computer system.
The mapping between the virtual addresses contained
in a user buffer and physical addresses used on the memory buses must be made, and remain, safely accessible
to the device. The virtual address mapping must be guaranteed to remain consistent between the device view and the
view of the kernel and processing cores for the duration of the device activity, and autonomous updating of the
relevant page mappings by the kernel must either be suppressed or correctly tolerated by the device hardware.

This consistent view of the translation can be established by either:
\begin{itemize}
\item using an IOMMU that interprets the  O/S maintained hardware page table in a similar way to the CPU cores, 
      generating page faults as required ; or
\item having the kernel \emph{lock} and \emph{pin} memory pages to freeze the translation between virtual and
      physical addresses within the virtual memory range. This translation must be provided to the device via a different mechanism, 
      and be frozen for at least the duration of the transaction.
\end{itemize}      

With the above conditions in place, it becomes possible for interconnect hardware to 
fetch and inject data directly from the user space process memory. 
This can even be exposed as \emph{remote memory access} where hardware mechanisms support
direct (and potenatially offloaded DMA) copy between user process memory
on two different computing nodes.
Such hardware remote memory access mechanisms are targeted in software by the Cray SHMEM and OpenSHMEM interfaces, 
and more recently in MPI-3 remote memory access primitives. Middleware supporting this
includes the Infiniband RDMA primitives and RDMA on Converged Ethernet (ROCE), and some of the
performance considerations discussed below may be fairly general. 

For the purposes of this work we focus on the 
\Intel Omni-Path Architecture (OPA), which has rapidly
acquired market share and is the basis of a number of large supercomputer installations.
One significant advantage of the CPU and interconnect technologies being provided by a single
vendor is the opportunity for \emph{integration}. Both the \XeonPhi and \Xeon lines have parts that
integrate \OPA interconnect on package.
The \XeonPhi 7250F is an example of a \KNL processor where Omni-Path is tightly integrated; both
discrete die for the CPU and a second die continaining \emph{two} PCIe x16 host fabric adaptors 
are combined in a single package. The \Xeon Gold 6126 is an \SKL that integrates an Omni-Path adaptor with
Skylake processor cores in a single package via a single on package x16 PCIe interface.

The driver software that interfaces between MPI and the Omni-Path hardware lives partly in user
space, and provides zero-copy DMA mechanisms with a driver interface (Performance Scaled Messaging 2) designed
to map well to higher level MPI primitives. 
The source code of the PSM2 stack is Open Source and available at 
\begin{center}
\href{https://github.com/01org/opa-psm2}{https://github.com/01org/opa-psm2}
\end{center}
and (along with the hfi driver in the Linux Kernel) may be consulted for further technical details. In
particular comments in \verb1 ptl_ips/ips_tidcache.h 1 may be useful.

This paper does not attempt to describe the software
and hardware implementation in detail, but rather treat the drivers as a black box.
We describe higher level software techniques that better exploit the devices from an empirical
basis, merely observing that certain strategies greatly enhance the measured performance.

In order to understand these results it suffices to understand only a few things. 
\begin{enumerate}
\item The \OPA device driver maintains a cache of translations; 
\item translation quanta must be both physically and virtually contiguous regions consisting of a power-of-two 
      number (1,2,4,\ldots 512) pages;
\item pinning memory and setting up translation requires a software overhead to be incurred for each such region
      a corresponding to a virtually contiguous buffer;
\item this overhead may be significant, and will be incurred for each such region;
\item when the Linux VM system is fragmented into small pages, this overhead may be paid as much as once \emph{per virtual memory page};
\item {\bf huge page memory pages (2MB) guarantee to suppress this overhead by up to a factor of 512 over default x86 pages (4KB);}
\item transparent huge pages may suppress this overhead when the virtual memory is \emph{not} already fragmented, but there are
      no mechanisms in Linux to \emph{guarantee} huge pages other than static reservation and special (mmap or shmget) allocation approaches.
\end{enumerate}

It is quite possible that the physical memory will contain many non-contiguous,
fragmented 4KB regions after a node has been running for some length of time. Dropping disk caches, ``compacting'' memory, and the Intel
Zone Sort kernel module are all presently recommended steps to address fragmentation and MCDRAM cache conflict issues with the Knight's Landing
processor and may help. Further, there is a standard assistance library \verb1 libhugetlbfs 1 that gives several mechanisms (some transparent) 
to assist code with obtaining huge pages.

However we document in the next subsection the approach taken in our study to \emph{guarantee} access to huge pages, since we found that transparent huge
pages were not, in practice, reliable. Guarantees are in many aspects of life preferable to optimistic probabilistic statements.
In a large supercomputer system, even a very modest per-node probability of poor performance will become
a stronger likelihood at large node counts, and steps that guarantee good performance are the best strategy.

\subsection{Cluster preparation for devolving huge page pool control}

Huge page pools may be statically allocated. It is impossible to select a single value for this
pool size that serves the needs of all applications since there are many and varied applications. 
We recommend to devolve huge page allocation sizing to users. 

This may be easily done using libhugetlbfs, since that captures most of the required administrative steps.
One possible approach is therefore to ask the adminstrators to configure each cluster node to support
libhugetlbfs, and \emph{devolve control of the huge page pool to the user owning the node at runtime} as follows:
\begin{verbatim}
sudo yum install libhugetlbfs-devel.x86_64
sudo yum install libhugetlbfs-utils.x86_64
sudo hugeadm --create-global-mounts

CFIL=/usr/bin/hugepage-reserve.c
XFIL=/usr/bin/hugepage-reserve

cat > $CFIL << EOF
#include <stdlib.h>
#include <stdio.h>
#include <unistd.h>

char command[1024];

int main(int argc,char **argv) 
{
  if (argc<2) exit(0);

  int num = atoi(argv[1]);

  if ( num<0    ) exit(0);
  if ( num>8000 ) exit(0);

  setuid(0);
  sprintf(command,"/usr/bin/hugeadm --pool-pages-min=2M:%d",num);
  printf("Executing command : %s\n",command); fflush(stdout);
  system(command);
  system("/usr/bin/hugeadm --pool-list");
}
EOF

cc $CFIL -o $XFIL
chmod 4755 $XFIL

\end{verbatim}

The above setuid binary will be executable by any user allowed to log in to the node
(typically controlled by the batch system) and performs range checks on the number of 2MB pages.
With the above code, the user allocated the computing resource can then dynamically resize the pool from zero to 16GB 
within their batch job, using the above setuid programme run on each cluster node 
to vet parameters and make the call to hugeadm:

\begin{center}
\verb1 mpirun -np N hugepage-reserve <PAGES> 1
\end{center}

Other approaches can of course be taken, consistent with this end goal. For example,
the pool size could be requested in job parameters and the resize performed in a
batch schedule prologue script.

It is already recommended by Intel to run ``drop cache''
and ``compact memory'' in the prologue scripts to maximise the chances of Linux Transparent Huge Pages (THP)
from using 2MB pages for large memory allocations.
\begin{verbatim}
   echo 3 > /proc/sys/vm/drop_caches
   echo 1 > /proc/sys/vm/compact_memory 
\end{verbatim}
However, these latter operations only enhance the probability of obtaining huge pages. Residual performance 
variability exists depending on the level of fragmentation into small 4KB regions of the virtual
memory system of a node, and only the use of explicit huge pages will guarantee huge page allocations.

\emph{Performance variability that lies beyond user control, and depends on the history of a node, leads
to a poor user experience. We believe it is important that the adminstrators on HPC sites using the \OPA 
interconnect take steps enabling hugepage allocations, similar to those used in this paper, to enable users to unlock the full
performance of \OPA in a reliable and predictable manner.}

\section{Halo Exchange PDE Communications Performance}

We investigate the performance of \OPA on \XeonPhi compute nodes on two test systems.

\subsection{Test systems}

Firstly on the Intel 'Diamond' cluster,  using dual rail \OPA  with 
\XeonPhi 7250 compute nodes. Secondly on the Cambridge University DiRAC cluster using
single rail \OPA and \XeonPhi 7210 compute nodes.
Finally, we also make use of dual socket \SKL Platinum 8170 nodes on the Intel 'Diamond' cluster with
single rail \OPA.

The peak interconnect bidirectional bandwidth for single rail is:
$$
100 \mathrm{Gbit/s} \times 2 \equiv 25 \mathrm{GB/s},
$$
and for dual rail the bidirectional line rate is,
$$
200 \mathrm{Gbit/s}  \times 2 \equiv 50 \mathrm{GB/s}.
$$
We would hope to come close to saturating this line rate on large packet sizes.

\subsection{Methodology}
The benchmark software uses the Grid QCD library \cite{Grid},
\begin{center}
\href{https://github.com/paboyle/Grid}{https://github.com/paboyle/Grid}
\end{center}
and the peformance benchmark \verb1 benchmarks/Benchmark_comms.cc 1 is used.

The tests are run on 16 nodes system is divided into a four dimensional cartesian communicator
with $2^4$ ranks, and one rank per node. These dimensions are called $\{ x,y,z,t\}$. 
Each node is given a four dimensional volume $L^4$, and
the global volume is $G^4 = (2L)^4$. The code mimics the communications patter for a halo exchange
PDE arising in quantum chromodynamics. 

Each node sends packets of size the surface $L^3$ to neigbours in each of $+x$, $-x$,
$+y$, $-y$,
$+z$, $-z$,
$+t$, and $-t$ directions, while concurrently receiving the neighbours data.

This is performed using \\
i) Buffers allocated with \verb1 memalign 1 and a single communication thread running 
   MPI\_Isend, MPI\_Irecv, MPI\_Wait sequences.\\
ii) Buffers allocated with explicit huge pages, and a single communication thread running 
   MPI\_Isend, MPI\_Irecv, MPI\_Wait sequences.\\
ii) Buffers allocated with explicit huge pages, and a multiple communication threads each running 
   MPI\_Isend, MPI\_Irecv, MPI\_Wait sequences on a distinct (duplicate) cartesian communicator.

The software has been modified to gain concurrency through the MPI stack consistent with the 
baseline requirements of the Intel MPI 2019 \emph{Technology Preview}.

\subsection{Memory allocation options}

For our default allocations we use the memalign routine:

\begin{verbatim}
#define GRID_ALLOC_ALIGN (2*1024*1024)
    CommBuf = (void *)memalign(GRID_ALLOC_ALIGN,bytes);
\end{verbatim}

After the system administrator support described above, the user code
sequence used to obtain explicit huge pages in a communications buffer is as follows:
\begin{verbatim}
  int mmap_flag = MAP_SHARED | MAP_ANONYMOUS | MAP_HUGETLB;
  CommBuf =(void *) mmap(NULL, MAX_MPI_SHM_BYTES, PROT_READ | PROT_WRITE, mmap_flag, -1, 0); 
\end{verbatim}

The full context of the call is in\\
\href{https://github.com/paboyle/Grid/blob/develop/lib/communicator/Communicator_base.cc}
     {https://github.com/paboyle/Grid/blob/develop/lib/communicator/Communicator\_base.cc}.

In a NUMA system, such as a dual socket \Xeon node, it is often convenient to place
communcations buffers in Unix shared memory, to enable ranks within the same node to use direct, multithreaded
access to buffers of neighbouring ranks on the same node but diffent NUMA domains.

\begin{verbatim}
#define GRID_SHM_PATH "/var/lib/hugetlbfs/group/wheel/pagesize-2MB/"
    sprintf(shm_name,GRID_SHM_PATH "/Grid_mpi3_shm_%d",rank);
    int fd=open(shm_name,O_RDWR|O_CREAT,0666);
    int mmap_flag = MAP_SHARED |MAP_POPULATE| MAP_HUGETLB;
    CommBuf = (void *) mmap(NULL, size_bytes, PROT_READ | PROT_WRITE, mmap_flag,fd, 0); 
\end{verbatim}

In this case an MPI rank can open the corresponding memory region of any other rank that lives on the same node
to accelerate intranode communication. The full context of the call is in\\
\href{https://github.com/paboyle/Grid/blob/develop/lib/communicator/Communicator_mpi3.cc}
     {https://github.com/paboyle/Grid/blob/develop/lib/communicator/Communicator\_mpi3.cc}.

\subsection{Hybrid MPI + OpenMP programming model}

The Grid code uses a hybrid OpenMP + MPI programming model. 

For programme execution within a single NUMA domain, thread parallelism gives a 
greater performance than MPI parallelism since exchanging messages involves redundant
memory system traffic than can be elimited using the single memory space available
to threading.

The code is very much NUMA aware, making use of MPI-3 features and also internally
developed support for shared memory. It is designed to use a hybrid of 
\begin{enumerate}
\item OpenMP threading (with pinned thread affinity) within a NUMA domain,
\item Unix shared memory interprocess communication between NUMA domains within a node
      (i.e. copies performed by \emph{all} threads and \emph{not} MPI transfers)
\item MPI messaging between nodes
\end{enumerate}

This gives the greatest single performance since redundant memory transfers are eliminated
within a NUMA domain; accesses across inter-socket connections like QPI or
UPI are suppressed by the surface to (per-socket) volume ratio; transfers over MPI are suppressed by 
the surface to (per-node) volume ratio which is even more favourable. MPI is \emph{not} explicitly
used for inter-socket, intra-node communication because it is better to use many threads to perform
these data copies than to use a single thread.

We find that Knight's Landing has a single socket and gives best single node performance using pure threading.
We have found that, particularly for Knight's Landing nodes, Intel MPI does not easily saturate wire
speed with a single MPI rank per node,  but does saturate with multiple ranks per node.
Thus, the model that gives best performance from a single node
execution point of view is thus placed in tension with the model that gives the best interconnect performance.
The solution has been to provide additional cores entering the networking stack concurrently from this a single
MPI process. 

This model has been enabled by a substantial effort first available in the Intel MPI 2019 Technology Preview, and a
multi-endpoint enhancement of the PSM2 library.
The MPI  and PSM2 messaging layers enable concurrency with the PSM2 library using a multiple endpoints from distinct
communicators. Since tag matching spaces are unique to each communicator, the use of distinct communicators on distinct
threads avoids thread contention for internal resource data structures.

We modified the relevant Communication objects in the code allocate additional, duplicate, Cartesian communicators,
to the desired level of thread concurrency.
\begin{verbatim}
  std::vector<MPI_Comm> communicator_halo;
  communicator_halo.resize (comms_threads);
  for(int i=0;i<comms_threads;i++){
    MPI_Comm_dup(communicator,&communicator_halo[i]);
  }
\end{verbatim}

These are then accessed by the corresponding OpenMP threads 0 \ldots comms\_threads-1, and used to 
advance the communications of packets in each direction concurrently through the networking stack. In particular
a runtime, command line selectable number (\verb1 --comms-threads <N> 1) of cores, where $N \in \{1,\ldots, 8\}$ 
are devoted to progressing communication even when communication and computation are overlapped in the code.

The differentiated behaviour of the multiple 
threads in a parallel region was controlled deterministically using
\begin{center}
 \verb1 omp_get_thread_num() 1.
\end{center}

\subsection{Results}

All results each table are obtained from a single run on the same set of nodes, and comparison between
different software strategies are obtained from the same set of Unix processes. Thus, the software
and hardware environments are maximally identical, and comparisons of different sofware code paths
and memory allocation strategies are ''like-for-like''.

Table~\ref{tab:2017_dual_rail} displays the performance obtained with cases i), ii), and iii)  using the Intel MPI
2017 library.

\begin{table}[hbt]
\begin{center}
\begin{tabular}{|c|c|c|c|c|c|}
\hline
Comms       & Seq    & Seq    & Concurrent & Concurrent &Threaded  \\
\hline
Pages       & huge   & memalign & memalign & huge  & huge      \\
\hline
Bytes       & \multicolumn{5}{c|}{2017.4 Intel driver bandwidth MB/s} \\ 
\hline
49152       &  1317.0  & 829.0  & 1641.2 &   384.3 & 399.8  \\
393216      &  5789.2  & 4702.9 & 8457.8 &  5995.7 & 5054.7  \\
1327104     &  8975.4  & 9392.6 &12104.2 &  9314.0 & 10585.3  \\
3145728     &  10731.8 & {\bf 434.8}  & {\bf 756.7} & 10739.2 &13984.5  \\
6144000     &  10970.9 & {\bf 750.5}  & {\bf 1209.6} & 11071.1 &13931.3  \\
10616832    &  11339.6 & {\bf 799.9}  &  980.0 & 11451.7 &12888.2  \\
16859136    &  11549.5 & {\bf 1201.7} & 1561.7 & 11607.9 &12272.8  \\
25165824    &  11740.0 & {\bf 1588.1} & 2007.4 & 11774.9 &12016.5  \\
\hline
\end{tabular}
\end{center}
\caption{\label{tab:2017_dual_rail}
We display the delivered bandwidth from DDR resident data on the 'Diamond' dual rail KNL cluster 
using Intel MPI 2017 and compiler suite with the Benchmark\_comms benchmark. 
In the column labelled ``Threaded'', eight communication threads are used.
Note that there is significant drop (bold entries) in delivered performance whenever explicit huges pages are not used.
}
\end{table}

In the ``Seq'' columns the x,y,z, and t communications are performed sequentially, one after the other,
while in the ``Concurrent'' columns asynchronous MPI calls are set pending for all directions concurrently.
In the ''Threaded'' column, a separate thread (and communicator) is used to concurrently enter MPI for each of
the eight directions.

Already it can be seen that there is a large qualitative difference between the use of explicit huge
pages in the buffer allocation and the use of standard Unix memory allocation mechanisms.
Note that there is significant drop (bold entries) in delivered performance whenever explicit huges pages are not used.
The drop does not appear with newly rebooted nodes, and is attributed to fragmentation of the  Linux virtual memory
system into 4KB regions. We find the performance results for the ``memalign'' allocation
are highly non-reproducible, and so a single run is taken as indicative of the ``typical'' problem.  The
results with explicit huge page allocations were completely reproducible and in many cases significantly better.

With the 2017 version of the Intel MPI library, and libpsm2 v10.2.235, there is no gain from concurrent threaded calls 
(column ``Threaded'') into the MPI stack from multiple threads, and in these
versions software prior to the 2019 technology preview, beta version and full release cycle a single
MPI rank on one node is not able to saturate the interconnect line rate of 50GB/s bidirectional bandwidth.

In Table~\ref{tab:2019_dual_rail}, we run using the Intel MPI and PSM2 2019 Technology Preview with the PSM2\_MULTI\_EP shell variable set. 
The additional concurrency enables up to 82\% of wirespeed
to be obtained from a single MPI rank on each node.
Note that there is significant drop (bold entries) in delivered performance whenever explicit huges pages are not used.

\begin{table}[hbt]
\begin{center}
\begin{tabular}{|c|c|c|c|c|c|}
\hline
Comms       & Seq    & Seq    & Concurrent & Concurrent &Threaded  \\
\hline
Pages       & huge   & memalign    & memalign        & huge  & huge      \\
\hline
Bytes       & \multicolumn{5}{c|}{New EP driver bandwidth MB/s} \\ 
\hline
49152       &  1429.8 &   372.7 &  1671.6 &    717.8 & 1543.9\\
393216      &  6096.1 &  4534.1 &  8705.0 &   5011.7 & 24013.2\\
1327104     &  9086.5 &  9000.9 & 13474.2 &   9093.8 & 31682.1\\
3145728     & 10654.8 & 10495.7 & 14892.0 &  10645.0 & 37645.2\\
6144000     & 11018.6 & 10577.1 & 14644.8 &  11107.1 & 40134.4\\
10616832    & 11488.4 & 11184.1 & 14299.4 &  11564.5 & 40786.4\\
16859136    & 11832.3 & {\bf  1116.5 }& {\bf  2055.3} &  11816.3 &  40698.8\\
25165824    & 12058.9 & {\bf 1843.3 }& {\bf 2521.9} &  11909.9 & 40636.4\\
\hline
\end{tabular}
\end{center}
\caption{\label{tab:2019_dual_rail}
We display the delivered bandwidth from DDR resident data on the 'Diamond' dual rail KNL cluster 
using Intel MPI 2019 and compiler suite with the Benchmark\_comms benchmark. 
Eight communication threads are used in the threaded column.
Note that there is significant drop (bold entries) in delivered performance whenever explicit huges pages are not used.
}
\end{table}

In Table~\ref{tab:2019_single_rail}, we run using the Intel MPI and PSM2 2019 Technology Preview with the PSM2\_MULTI\_EP shell variable set. 
The additional concurrency enables up to 89\% of wirespeed
to be obtained from a single MPI rank on each node.
Note that there is significant drop (bold entries) in delivered performance whenever explicit huges pages are not used.

For comparison, we also make use of dual socket \SKL Platinum 8170 nodes on the Intel 'Diamond' cluster with
single rail \OPA, and display the results in Table~\ref{tab:2019_skl_single_rail}. Performance remains more
reproducible and better with huge page allocations; however the losses are certainly not as dramatic as
arises with the much slower Knight's Landing CPU cores in the above set of results. The effects can still
be significant, and use of \emph{libhugetlbfs} in production \Xeon clusters may continue to make sense for
site administrators, but is perhaps less important than for systems using \KNL computing nodes. Since the
effects are not typically reproducible, and depend on a history of VM fragmentation, it is, of course, hard to 
establish the worst case.

\begin{table}[hbt]
\begin{center}
\begin{tabular}{|c|c|c|c|c|c|}
\hline
Comms       & Seq    & Seq    & Concurrent & Concurrent &Threaded  \\
\hline
Pages       & huge   & memalign    & memalign        & huge  & huge      \\
\hline
Bytes       & \multicolumn{5}{c|}{EP driver bandwidth MB/s} \\ 
\hline
49152       &   1296.8  & 390.8 & 1620.9  & 482.9 & 756.3 \\
393216      &   4427.7  & 3856.8& 8401.7  & 4062.1&11504.5\\
1327104     &   7501.0  & {\bf 577.8} & 11702.4 & 7953.5&18924.5\\
3145728     &   8636.9  & {\bf 830.8} & 12901.3 & 8811.9 &21776.1\\
6144000     &   9279.1  & {\bf 611.9} & 13154.4 & 9159.0&21812.5\\
10616832    &   9574.6  & {\bf 474.5} & 12580.4 & 9642.5&21927.8\\
16859136    &   9632.2  & {\bf 547.0} & 8956.7  & 9620.5&21880.2\\
25165824    &   10137.6 & {\bf 716.0} & {\bf 1433.0}  & 9971.9 &22281.0\\
\hline
\end{tabular}
\end{center}
\caption{\label{tab:2019_single_rail}
We display the delivered bandwidth from DDR resident data on the 'DiRAC' single rail KNL cluster 
using Intel MPI 2019 and compiler suite with the Benchmark\_comms benchmark. 
Eight communication threads are used in the threaded column.
Note that there is significant drop (bold entries) in delivered performance whenever explicit huges pages are not used.
}
\end{table}

\begin{table}[hbt]
\begin{center}
\begin{tabular}{|c|c|c|c|c|c|}
\hline
Comms       & Seq    & Seq    \\
\hline
Pages       & huge   & memalign   \\
\hline
Bytes       & \multicolumn{2}{c|}{EP driver bandwidth MB/s} \\ 
\hline
49152       &   7651.6   & 7397.5  \\
393216      &   15619.7  & 13776.5 \\
1327104     &   17096.6  & 16031.6 \\
3145728     &   18928.2  & 18463.4 \\
6144000     &   20043.6  & 12699.8 \\
10616832    &   21710.1  & 15457.0 \\
16859136    &   21813.0  & 17238.9 \\
25165824    &   22796.9  & 19418.4 \\
\hline
\end{tabular}
\end{center}
\caption{\label{tab:2019_skl_single_rail}
We display the delivered bandwidth from the 'Diamond' single rail Skylake cluster.
using Intel MPI 2019 and compiler suite with the Benchmark\_comms benchmark. 
There is some evidence of a modest performance degradation from regular memory compared
with the use of huge pages, but the effect is certainly not as dramatic as arises
with the much slower Knight's Landing CPU cores in the preceeding tables.
}
\end{table}

\FloatBarrier

Tables~\ref{tab:perf_single_rail} and \ref{tab:perf_dual_rail} show the PDE
sparse matrix multiply performance delivered on the two test systems 'Diamond' and
'DiRAC'. After the (substantial) tuning work discussed in this paper
the perfromance gain over the original performance is more than a factor of two.
Further, after these modifications, the second rail can add as much as 30\% to the 
total application throughput. 

\begin{table}[hbt]
\begin{center}
\begin{tabular}{|c|c|c|c|c|c|}
\hline
Local vol & $D_w$ & $D_w^{eo}$  & $sD_w$ & $sD_w^{eo}$ \\
\hline
$12^4$    & 153595& 127253 & 218260 & 168000 \\
$16^4$    & 207006& 196695 & 326736 & 300858 \\
$24^4$    & 276629& 262282 & 417697 & 417786 \\
\hline
\end{tabular}
\end{center}
\caption{\label{tab:perf_single_rail}
We display the delivered (single precision) application performance for multi-node 
sparse matrix application, for various sparse matrices, on the 'DiRAC' single 
rail cluster versus the problem size per node.
}
\end{table}

\begin{table}[hbt]
\begin{center}
\begin{tabular}{|c|c|c|c|c|c|}
\hline
Local vol & $D_w$ & $D_w^{eo}$  & $sD_w$ & $sD_w^{eo}$ \\
\hline
$12^4$    & 188241 & 153606 & 290627 & 234696 \\
$16^4$    & 286771 & 249671 & 416346 & 380685 \\
$24^4$    & 350112 & 341795 & 495970 & 502127 \\
\hline
\end{tabular}
\end{center}
\caption{\label{tab:perf_dual_rail}
We display the delivered (single precision) application performance for multi-node 
sparse matrix application, for various sparse matrices, on the 'Diamond' dual rail cluster versus the problem size
per node. 
}
\end{table}


\section{Performance of synchronous gradient descent vector reduction}

Machine learning is currently a high profile and emerging field. There are two distinct
activities in the application of neural networks to complex problems. 
Firstly a network must be trained, ideally rapidly, against a set of reference data or some measure
of ``correctness'' of the network output. This is by design a non-linear optimisation problem
against a very large set of network weights or coefficients.

The simplest gradient descent training algorithm would introduce a cost function, 
$$\chi_i(\vec{p}),$$
giving a measure of the differences (e.g. mean square deviation) between the output of the
network and a reference desired value on each sample $i$ of the training data.
The most naive algorithm would be a \emph{steepest descent} iterating

$$\vec{p} \to \vec{p} - \alpha \sum\limits_i  \vec{\nabla}_p \chi_i(\vec{p}).$$

Such an algorithm moves inefficiently through phase space, leading to slow convergence and also often leads to \emph{overfitting}
(a condition where the high dimensional network is trained to ``remember'' the specific training
cases but not respond appropriately to \emph{new} data). Stochastic Gradient Descent (SGD) replaces the above sum with a randomly
selected subsample of size $N$ (a mini-batch) of the data.

$$
\vec{p} \to \vec{p} - \alpha \sum\limits_{i \in \{ \mathrm{Random}~\mathrm{Batch }\}}  \vec{\nabla}_p \chi_i(\vec{p}).
$$

\begin{figure}[hbt]
\includegraphics[width=0.6\textwidth]{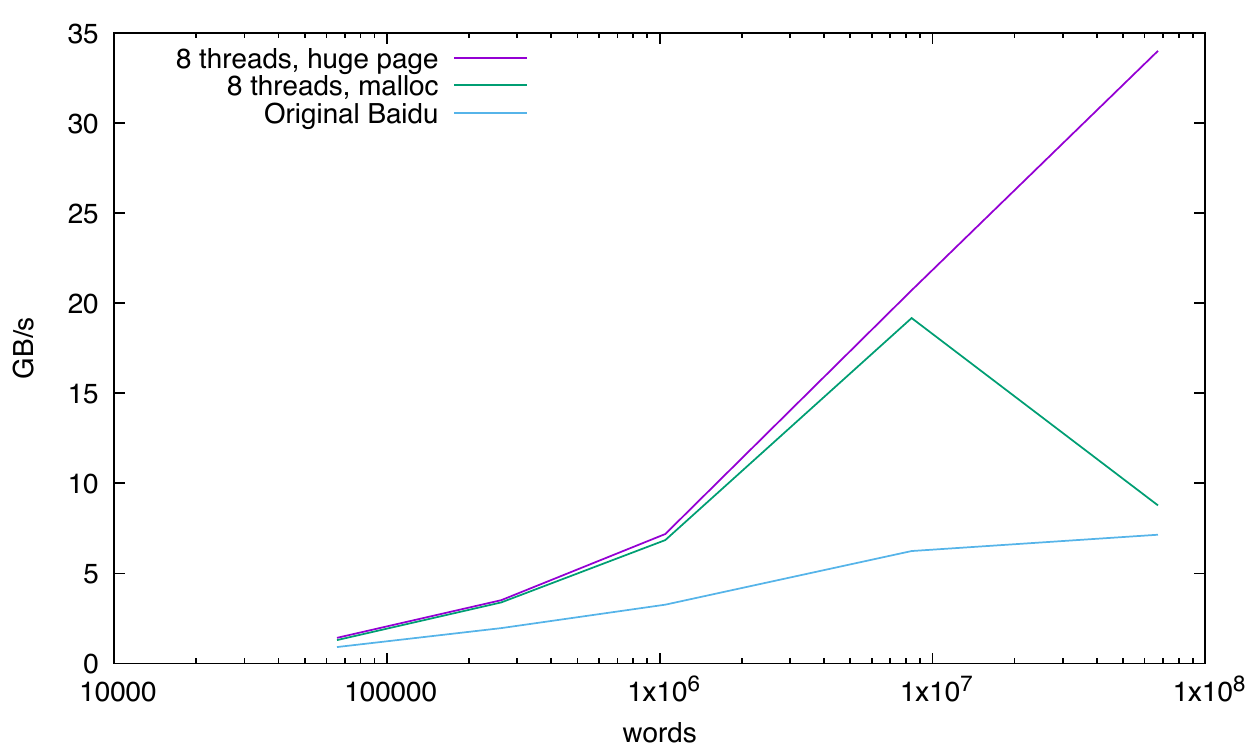}
\caption{\label{fig:Bandwidth} Bandwidth delivered before and after optimisation steps were taken. Only the (blocking) MPI
calls themselves were included in the timing. In the optimised code 35GB/s bidirectional bandwidth is delivered, and 
that is 70\% of line-rate, and contrasts
well against the 10\% delivered by the original code.}
\end{figure}

\begin{figure}[hbt]
\includegraphics[width=0.6\textwidth]{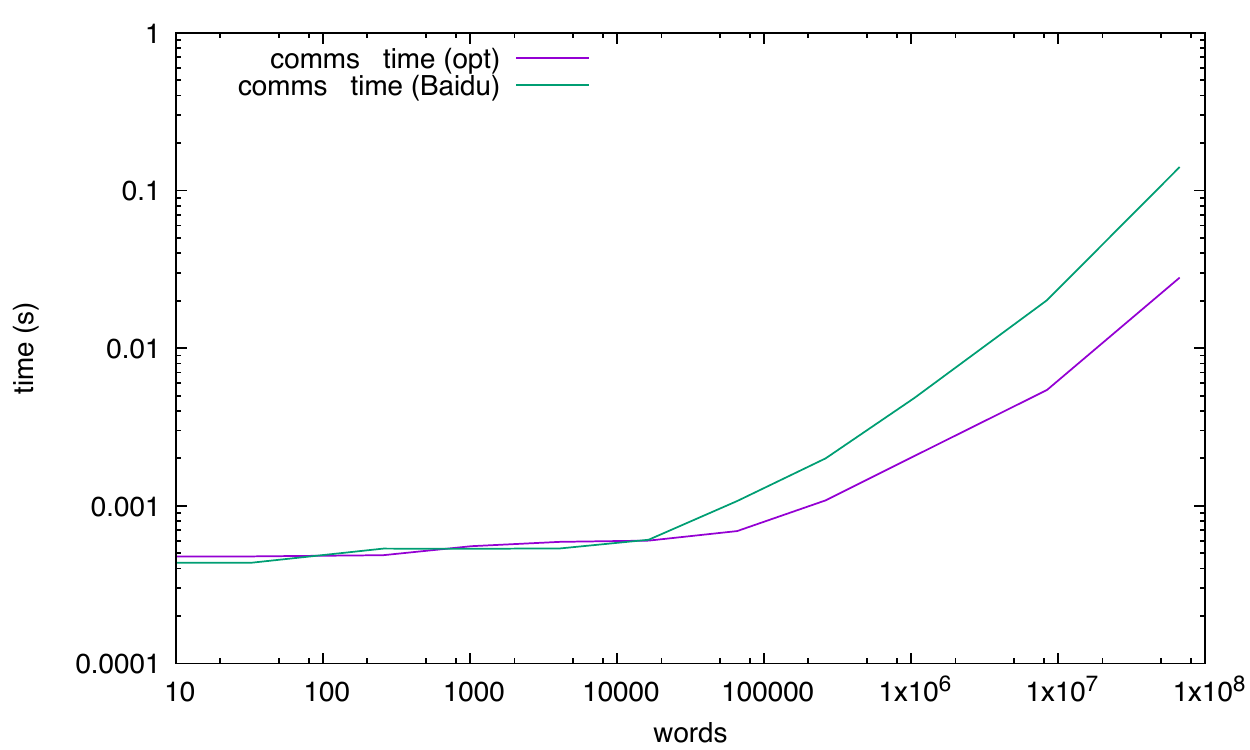}
\caption{\label{fig:time_comms} 
Total time for the communication vs. vector length before and after our optimisation.
Latency dominates for small vectors, but a substantial gain 
is possible in the large packet/bandwidth limited end of the curve. Note the scale
is logarithmic.
}
\end{figure}

\begin{figure}[hbt]
\includegraphics[width=0.6\textwidth]{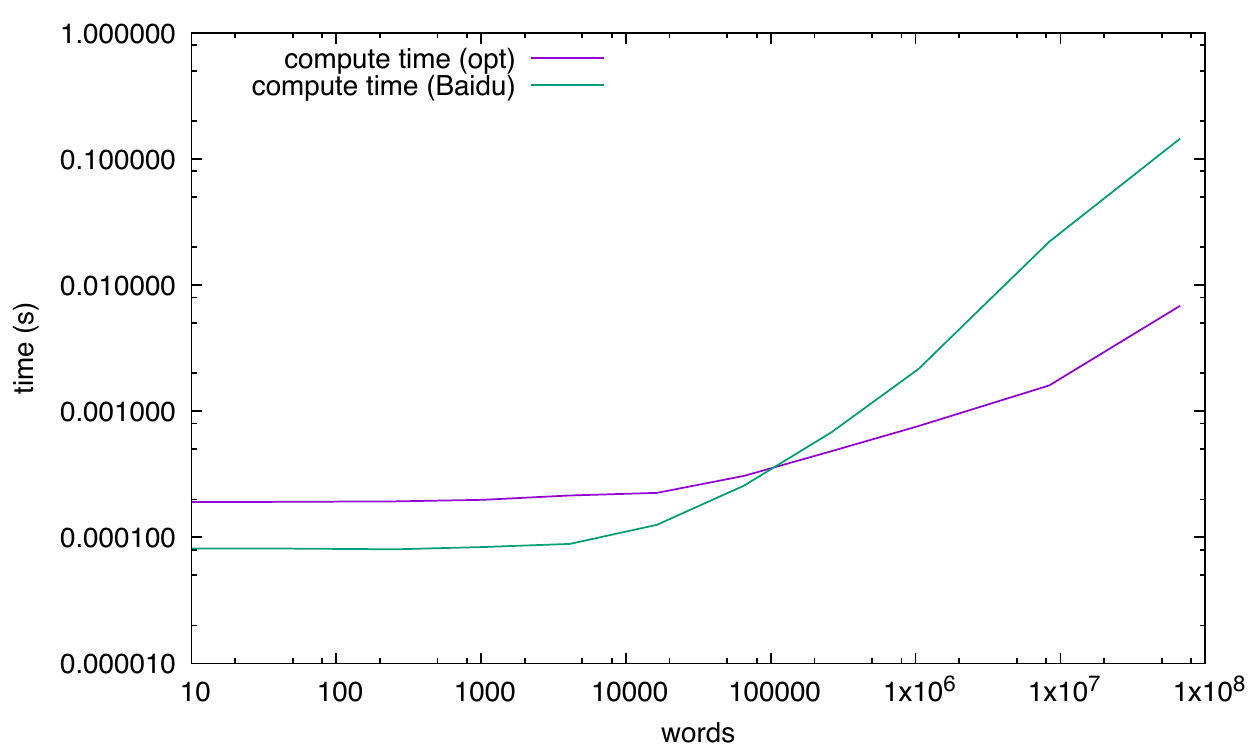}
\caption{\label{fig:time_compute} 
Total time for all computation, buffer copying and memory allocation vs. vector length before and after our optimisation.
Threading overhead dominates the ``optimised'' code on small vector lengths with 64 active threads.
Clearly, further optimisation is possible,  making the thead count used vary with the vector
length would yield a solution that is always optimal for all vector lengths. This study
does not attempt to do this and focuses on the large vector performance since this is 
most relevant for large, complex and demanding neural network problems.
}
\end{figure}

\begin{figure}[hbt]
\includegraphics[width=0.6\textwidth]{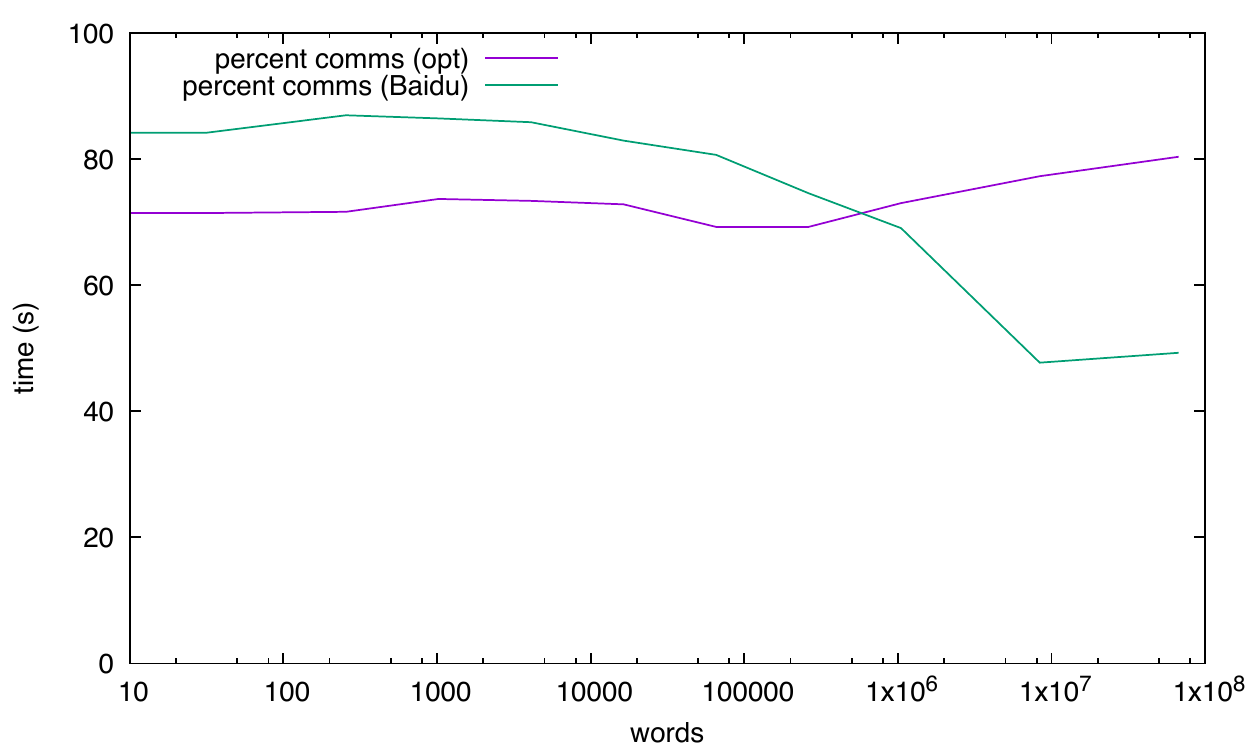}
\caption{\label{fig:percent_comms}
Percentage of time spent in communications calls vs. vector length before and after our optimisation.
The computation becomes dominant at large vector lengths in the original code, but is sub-dominant
in the optimised code \emph{despite} the optimised code being over 10x faster. This is not unreasonable
since the threading of the relevant loops should gain a factor of O(64) on many core processors.
}
\end{figure}

\begin{figure}[hbt]
\includegraphics[width=0.6\textwidth]{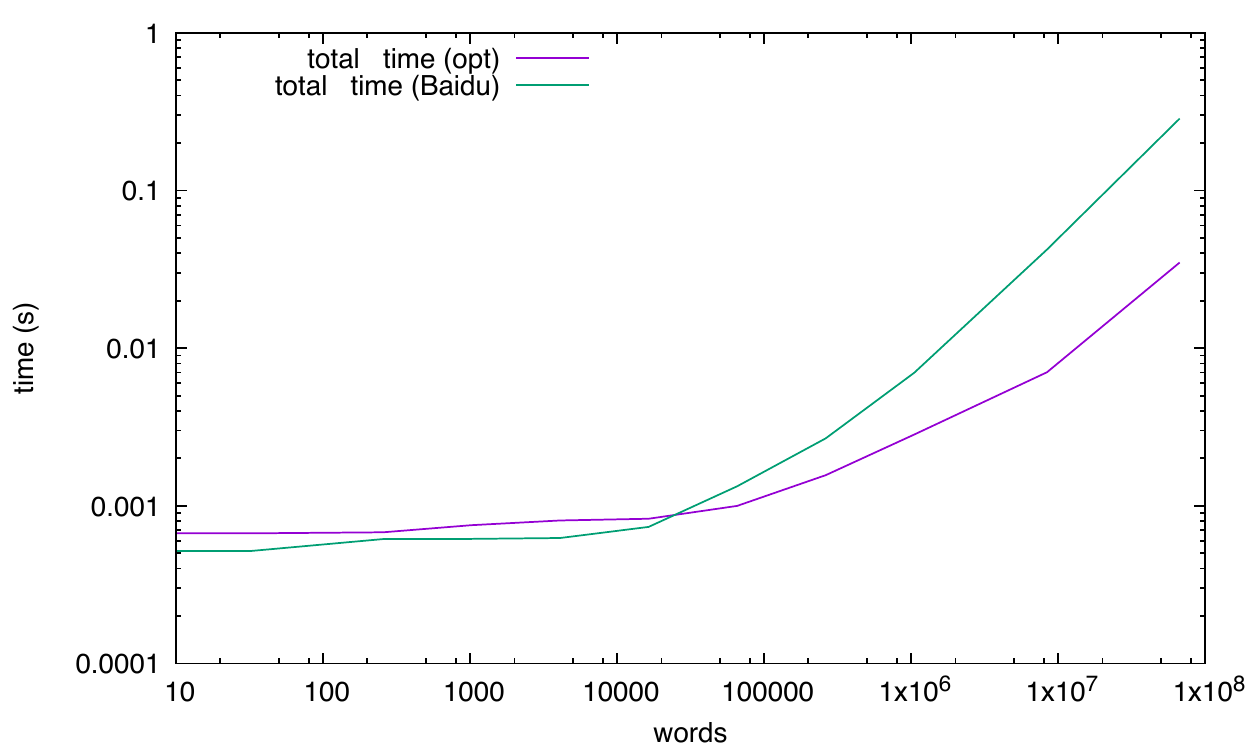}
\caption{\label{fig:time_total}
Wall clock time per reduction call vs. vector length before and after our optimisation.
The large vector reduction performance is ten times better after our optimisations on
large vector lengths. The gain includes both computation acceleration and communication acceleration.
}
\end{figure}

The distribution over MPI of the SGD is a key to the 
acceleration of the training of Neural Networks, and the single biggest to to faster ``Learning''
in the high profile field of Machine Learning.
If the elements of the mini-batch are shared among computing nodes, the distribution is called data parallelism, and
if the entire gradient in the above summation is accumulated across nodes, summed globally and the parameters updated
the algorithm is known as \emph{synchronous} SGD.

Since the aim is to engage as many nodes as possible to concurrently work on training neural networks, we aim
to subdivide the minibatch of fixed size $N$ onto $P$ processing nodes. There will be $N/P$ samples per node and
as $P$ is increased, the number of samples per node falls. This limit is known as the strong scaling limit, where
the global volume of work stays fixed, and the work per node \emph{decreases} as the number of nodes is increased.
In the strong scaling limit, the dominance of communications is likely as there will eventually be very
few elements of the batch processed by each node ($N/P \to 1$), while there a fixed, large number of parameters 
whose gradients must be reduced.

We now consider how to accelerate the  Baidu Research ``allreduce'' optimised reduction code\cite{Baidu}
\begin{center}
\href{http://research.baidu.com/bringing-hpc-techniques-deep-learning/}
     {http://research.baidu.com/bringing-hpc-techniques-deep-learning/}
\end{center}
available on GitHub as open source code,
\begin{center}
\href{https://github.com/baidu-research/baidu-allreduce}
     {https://github.com/baidu-research/baidu-allreduce}
\end{center}
as the basis for our study of acceleration under \OPA. This code is a good example of the 
important task of optimising the performance limiting steps in distributed machine learning.

We use the same 'Diamond' cluster as a test system, and demonstrate how to get near wirespeeed performance from 
dual rail \OPA cluster nodes. In order to obtain this speed up several modifications to the code were again required,
which we document here.

{\bf Improved memory allocation:}
Firstly, the original interface returns a new vector buffer allocated with the ``C++ new'' operator. It is
the responsibility of the caller routine to delete this vector. This unfortunately dictates that an entire
vector \emph{must} be allocated and deallocated for \emph{every} call to the reduction operation.
Tying memory management to the operations is not the optimal design, since hot loops may be best programmed
with reuse of all memory allocations.

Further, the benchmark code ties the accelerated reduction software interface to a specific memory allocation/deallocation interface:
the CPU and GPU version of the timing loop both call different deallocation mechanisms, which provides an inconsistent
software interface that depends on the compilation target. In both aspects the coupling of the interface to memory management is not ideal.

Rather, it is likely that method dealloc and alloc routine provided by collectives.h should have been used
consistently, with use of any other ``free'' operation declared illegal.
Further, this would have enabled simpler modification of the allocation and deallocation implementation.
It would also have been better to \emph{separate} the vector \emph{allocation} from the reduction \emph{operation}; so that in hot loops
the programmer could choose to reuse the same allocation.

Our first optimisations were high level: i) remove the expectation that the caller deallocates the returned vector; ii) the reduction
routine was modified to \emph{cache} the most recently returned pointer and the vector length for reuse
if a subsequent call does not require a longer vector. Deallocation/reallocation is suppressed to 
points in the code where the vector length increases. This change was made by introducing an caching ``Alloc'' and ``Dealloc'' routine
for internal vectors as a quick and dirty solution.
\footnote{It is worth commenting that in the Grid code above, a 
STL compliant C++ allocator is used consistently in the code, using template typedefs, to replace the standard allocator.
The ten most recently deallocated large vector allocations are cached, with a round robin victim and lazy release, so that repeated
reallocation of same sized vectors is efficient. This avoids releasing memory to the operating system in tight inner loops. Memory allocation 
caching is a valid and useful HPC optimisation and can be made generally applicable. It is even relatively simple if consistently
using STL vectors and template typedef's.
}
\begin{verbatim}
static float *buffer;
static float *output;
static size_t allocated_length; 
void Alloc(size_t length)
{
  if ( length == 0 ) length=1;
  if ( length > allocated_length ) {
    if ( allocated_length > 0 ) { 
      Dealloc();
    } 
    buffer = alloc(length);
    output = alloc(length);
    allocated_length=length;
  }
}
void Dealloc(void)
{
  dealloc(buffer);
  dealloc(output);
  allocated_length=0;
}
\end{verbatim}

Subsequently, we were only then able to iii) replace the allocation with a huge page mmap call.
\begin{verbatim}
std::map<float *,size_t> allocations;

// Allocate a new memory buffer on CPU or GPU.
float* alloc(size_t size) {

  // Align up to huge page boundary
  const uint64_t TwoMB = 2*1024*1024;
  size = size * sizeof(float);
  size = (size + (TwoMB ) ) & (~(TwoMB-1)); 

  float *buf =(float *) mmap(NULL, size, PROT_READ | PROT_WRITE,  
                             MAP_HUGETLB|MAP_SHARED| MAP_ANONYMOUS, -1, 0); 
  assert (buf!=MAP_FAILED);
  bzero(buf,size);            // first touch

  allocations[buf]=size;      // track allocations
  return buf;
}
// Deallocate an allocated memory buffer.
void dealloc(float* buffer) {
  size_t size = allocations[buffer];
  munmap(buffer,size);
  allocations.erase(buffer);
}
\end{verbatim}

It would be best to redesign the software interface to a) exclusively allocator and deallocator routines that
may then be changed in implementation specific optimisations, and b) allow the user to preallocate the return vector which could
then be reused multiple times in performance critical code.

Irrespective of how the implementation changes as a long term solution, we have demonstrated that the code would be improved by
i) allowing alternate allocators to be used, including explicit huge page allocators;
ii) decoupling the allocation and deletion of buffers from the operation of reduction.
This will allow specialised methods to obtain memory to be used, and to amortize allocation overhead (including
the mapping of pages into a process) across multiple reduction calls.

{\bf Threaded communications calls:} The communicator was duplicated a variable number of ways (eight ways in our
reported results), and the same number of communications threads were implemented, introducing a separate communicator
for each to provide more concurrent entrancy into the \verb1 MPIGlobalState 1 communication object . 
\begin{verbatim}
   int comm_threads;
   std::vector<MPI_Comm> comm_duplicates;
\end{verbatim}
These were initialised appropriately,
\begin{verbatim}
  mpi_error = MPI_Init_thread(NULL,NULL,MPI_THREAD_MULTIPLE,&provided);
  assert (provided == MPI_THREAD_MULTIPLE);

  global_state.comm_threads=threads;                          
  global_state.dups.resize(threads);                          
  for(int t=0;t<global_state.comm_threads;t++){               
    MPI_Comm_dup(MPI_COMM_WORLD,&global_state.dups[t]);       
  }                                                           
\end{verbatim}
and then available to use with thread concurrency. We wrote a quick wrapper
around MPI\_Sendrecv to break the work into multiple, threaded transfers.
\begin{verbatim}
static void GetWork(int nwork, int me, int & mywork, int & myoff,int units){
  int basework = nwork/units;
  int backfill = units-(nwork%units);
  if ( me >= units ) { 
    mywork = myoff = 0;
  } else { 
    mywork = (nwork+me)/units;
    myoff  = basework * me;
    if ( me > backfill ) 
        myoff+= (me-backfill);
  }
  return;
};
void OMP_Sendrecv_float(const float *sendbuf, int sendcount, int dest  , int sendtag,
                              float *recvbuf, int recvcount, int source, int recvtag)
{
#pragma omp parallel 
  {
    int t = omp_get_thread_num();
    if ( t < global_state.comm_threads ) {
      MPI_Comm comm = global_state.dups[t];
      int swork, rwork, soff, roff;
      GetWork(sendcount,t,swork,soff,global_state.comm_threads);
      GetWork(recvcount,t,rwork,roff,global_state.comm_threads);

      MPI_Sendrecv((const void *)&sendbuf[soff],swork,MPI_FLOAT, dest  , sendtag,
                   (void *)      &recvbuf[roff],rwork,MPI_FLOAT, source, recvtag,
                   comm, MPI_STATUS_IGNORE);
    }
  }
}
\end{verbatim}

The concurrency in the MPI stack is particularly important for dual rail systems.

{\bf Threaded copy and computation:}
Finally, the original loops running on CPU cores had no multi-core concurrency and were consequently inefficient.
The buffer copies and local vector arithmetic addition were (thread) parallelised using OpenMP parallel for loops.
This accelerates the non-communication portions of the code and is particularly relevant for large vector length reductions.
\begin{verbatim}

void copy(float* dst, float* src, size_t size) {
#pragma omp parallel for
  for(size_t i = 0; i < size; i++) {
    dst[i]=src[i];
  }
}

void reduce(float* dst, float* src, size_t size) {
#pragma omp parallel for
  for(size_t i = 0; i < size; i++) {
    dst[i] += src[i];
  }
}
\end{verbatim}

The bandwidth delivered by the code with various optimisations in place is shown in Figure~\ref{fig:Bandwidth}.
Only the MPI calls themselves are included in the timing here, so that memory allocation, deallocation, local computation
and copying are \emph{not} included.
The comparisons are quite interesting; the bidirectional bandwidth can be raised from just over 5GB/s to 35GB/s using the
combination of the improved huge page memory and concurrent communicating threads. We emphasize
that the peak is 50GB/s bidirectional, and that we are approaching this peak. If huge pages are not used, odd drops
for large vector performance can be seen. Since the network peak is substantially below both the DDR and the MCDRAM peak performance
there is not much difference between the two NUMA domains available in flat mode. Overall, a seven fold improvement in delivered
performance is possible with relatively minor changes to the code.

The communication time is measured as solely the time spent in Sendrecv calls.
This is displayed, both before and after the introduction of concurrency in the communication
calls in Figure~\ref{fig:time_comms}. This shows that a substantial gain is afforded, as is to be expected
from the above bandwidth plot.

We now consider, Figure~\ref{fig:time_compute} which shows the time for the \emph{non}-MPI overhead 
of memory allocation, deallocation, local computation and copying.
Ideally, the overheads would be small; we can see that the threading of loops,
and elimination of allocation overhead has given a substantial gain in the non-communication
elements of the reduction code.

We display the percentage of time spent communicating in Figure~\ref{fig:percent_comms}. Ideally 
the problem should be communications limited, and the communications should proceed are wirespeed.
This is clearly not the case for the original code. We can see that in the original code the overhead becomes large
for long vector reduction, and in fact drops to below 50\%. In contrast, despite delivering 7x greater bandwidth
on the wires, the optimised code also become communication \emph{dominated} at roughly 80\% communications in the
limit of large vectors.

Finally, we display the wall clock time per reduction call vs. vector length before and after our optimisation, in
Figure~\ref{fig:time_total}. The large vector reduction performance is ten times better after our optimisations on
large vector lengths. In the limit of strong scaling synchronous machine learning, this is a direct ten fold improvement in the
rate limiting step. In this code 35GB/s bidirectional bandwidth is delivered, and that is 70\% of the 50GB/s bidirectional linerate, and contrasts
well against the 10\% delivered by the original code. Other (non-communication) elements of the code are also improved with our modified
code, and the total speed is around 10x for large vectors.

\section{Discussion and conclusions}

We have demonstrated how to obtain near line-rate bidirectional performance from two very different classes of HPC
code on \OPA with a single MPI rank operating per node, suitable to hybrid OpenMP and MPI programming. 

Several changes to common practice were required:
\begin{itemize}
\item Administrator support to devolve permission to users to reserve and allocate huge pages on compute nodes that
      have already been allocated to the users by the batch scheduling system.
\item Custom allocation of communications buffers using Unix mmap calls with MAP\_HUGETLB flags. This may be performed
      either using anonymous maps when there is one MPI rank per node, or using the hugetlbfs filesystem to share memory
      between ranks in a multi-socket node.
\item Duplicating communicators so that multiple communications threads can be allocated by the programmer under OpenMP to
      entering the MPI stack on distinct communicators.
\end{itemize}

The optimisations here enable both Quantum Chromodynamics codes and Machine Learning codes to saturate between 80\% and 90\%
of bidirectional wirespeed performance on 16 nodes giving between four and ten times delivered network performance gains over the original
codes. Further, performance anomalies associated with fragmented small page virtual memory systems are completely avoided.

The modified version of the Baidu reduction code used in these benchmarks is available at:
\begin{center}
\href{https://github.com/paboyle/baidu-allreduce}{https://github.com/paboyle/baidu-allreduce}
\end{center}

The code is not ready for immediate use, and rather an interface redesign is required incorporating the
above considerations. Once the interface has been updated, there is the potential for up to ten fold acceleration of
synchronous stochastic gradient descent, in the strong scaling limite,
needed for massively distributed (and massively accelerated) machine learning. Near wirespeed bidirectional performance
is obtained on dual rail \OPA showing that after these suggested modifications are made, the code is very hard to further improve, as hard engineering limits
are being approached.

Finally, we conclude by repeating an earlier comment: 
\emph{Performance variability that lies beyond user control, and depends on the history of a node, leads
to a poor user experience: we believe it is important that the adminstrators on HPC sites using the \OPA 
interconnect take steps enabling hugepage allocations, to enable users to unlock the full
performance of \OPA in a reliable and predictable manner.}

{\bf Acknowledgements:}  
This gradient reduction optimisation in this work has been performed under 
the Alan Turing Insitite (ATI)-Intel strategic partnership.
P.B. is an Alan Turing Institute Fellow, and acknowledges Royal Society Wolfson Research Merit Award WM160035, and
STFC Grants ST/K005790/1, ST/K005790/1, ST/P002447/1, and ST/P000630/1. 
G.C. acknowledges funding by Intel and an STFC IAA award, and
is supported by STFC, grant ST/L000458/1 and ST/P002447/1. University of Edinburgh has received funding as an Intel Parallel Computing
Centre. C.L. acknowledges support through a DOE Office of Science Early Career Award and by US DOE Contract DESC0012704(BNL).
We wish to thank Camilo Moreno, Karthik Raman, and Joe Curley at Intel, and Norman Christ at Columbia for useful
conversations and advice. 
We wish to thank James Southern, Eng Lim Goh, Daniel Faraj at SGI for useful conversations, and the provision
of benchmarks on SGI ICE-XA systems.
We wish to thank James Erwin, and Edward Mascarenhas at Intel for 
developing and providing the Intel MPI 2019 technology preview. We thank staff at Intel, Brookhaven Laboratory, 
Stuart Rankin and Cambridge University for making system configuration changes that enabled this 
huge page study to be performed.
We thank Intel for access to the Diamond cluster.
We thank STFC, Cambridge University, and DiRAC project dp008 for access to the DiRAC KNL cluster at Cambridge.
We thank Antonio Rago and Plymouth University for temporary root access to their Broadwell OPA cluster.
We thank Brookhaven National Laboratory Computing Science Initiative for access to the Brookhaven institutional cluster.

\end{document}